\documentclass[]{spie}

\usepackage[utf8]{inputenc}
\usepackage{amsmath,amsfonts,amssymb}
\usepackage{graphicx}

\usepackage[colorlinks=true, allcolors=blue]{hyperref}
\usepackage{multirow}
\usepackage{booktabs}

\title{Wavefront Profiling via correlation of GLAO open loop telemetry}

\author[a]{Eden A. McEwen}
\author[b]{Ryan Dungee}
\author[b]{Mark Chun}
\author[a]{Jessica R. Lu}
\author[c]{Olivier Lai}
\author[b]{Christoph Baranec}
\affil[a]{Physics Department, University of California, Berkeley, CA 94720, USA}
\affil[b]{Institute for Astronomy, University of Hawai'i at M\={a}noa, Hilo, HI 96720-2700, USA}
\affil[c]{UCA, Observatoire de la C\^ote d'Azur, CNRS, Laboratoire Lagrange, France}
\authorinfo{Further author information: (Send correspondence to E.A.M.)\\E.A.M.: E-mail: emcewen@berkeley.edu 
}

\begin{document}
\maketitle

\begin{abstract}
    Adaptive Optics (AO) used in ground based observatories can be strengthened in both design and algorithms by a more detailed understanding of the atmosphere they seek to correct. Nowhere is this more true than on Maunakea, where a clearer profile of the atmosphere informs AO system development from the small separations of Extreme AO (ExAO) to the wide field Ground Layer AO (GLAO). Employing telemetry obtained from the  'imaka GLAO demonstrator on the University of Hawaii 2.2-meter telescope, we apply a wind profiling method that identifies turbulent layer velocities through spatial-temporal cross correlations of multiple wavefront sensors (WFSs). We compare the derived layer velocities with nearby wind anemometer data and meteorological model predictions of the upper wind speeds and discuss similarities and differences. The strengths and limitations of this profiling method are evaluated through successful recovery of injected, simulated layers into real telemetry. We detail the profilers' results, including the percentage of data with viable estimates, on four characteristic  'imaka observing runs on open loop telemetry throughout both winter and summer targets. We report on how similar layers are to external measures, the confidence of these results, and the potential for future use of this technique on other multi conjugate AO systems.
\end{abstract}

\section{INTRODUCTION}
    \label{sec:intro}  

    
    Adaptive optics (AO) has been used in the last few decades to correct Earth's atmospheric distortions and provide nearly diffraction-limited images for the world's largest ground based telescopes. Traditional, single conjugate AO systems take wavefront measurements from a guide star and correct path length differences of a single optical path with a deformable mirror (DM). Ground Layer AO (GLAO), which is designed for a partial correction across a wider field of view (FOV), combines wavefront measurements from multiple WFS spread out over the field to correct for common turbulence at lower atmospheres\cite{Rigaut2002glao}. Averaging over multiple WFSs intentionally gives an imperfect correction; prioritizing the correction of the lower layers of atmospheric turbulence ideally averages out the upper atmosphere and provides a more general correction over a wider FOV. 
    
    The 'imaka instrument has implemented GLAO on the University of Hawaii's 2.2-meter telescope on Maunakea observing with up to five natural guide star WFSs distributed over a 24'$\times$18' field of view and a science camera covering the center 11'x11' in the B, V, R, and I-bands \cite{chun2016imaka}. Typical guide star constellations include 3-5 stars in an asterism with diameter of about 14-18 arcminutes. 'imaka collects both focal plane images and AO telemetry data (e.g. WFS slopes and DM voltages) when the AO system is running in both open- and closed-loop configurations.  Data in a number of observing campaigns has been collected over the last 5 years. A night of data typically includes 20-150 AO circular buffers, 30 to 150-seconds long,  consisting of up to 5 WFS, each returning an 8x8 grid of wavefront sensor centroids/slopes and deformable mirror voltages at a sampling rate of around 180 Hz.
    
    In this study, we use open loop telemetry data as a window into the motion and evolution of the turbulence as it moves across the telescope's line of sight. Spatially correlating each WFS with itself, or auto-correlations, at increasing separation in time between frames highlight correlation peaks that move with at the turbulence layer velocities somewhere along the line of sight to the guide star \cite{wilson2002SLODAR}. Correlating WFSs with other WFSs, or cross-correlations, highlight correlation peaks that move within the volume of atmosphere shared by the two sensors. In other words, the auto-correlations show moving turbulence at all altitudes, while the cross correlations show turbulence moving at lower heights. 
    
    Previous work has shown wind profiling generating useful insights for site testing and AO optimization \cite{cortes2013analysis, wang2008atmospheric}. Extending wind profiling with automated detection \cite{guesalaga2014cn2} to  identify upper atmospheric layers could make the method useful for GLAO control schemes. Here we seek to test the strengths and limitations of the method on Maunakea, specifically comparing to externally measured wind speeds and altitudes. Implementing automatic detection and summarizing over many telemetry files provides further insight into its applciation and uses. 
    
    This paper will first discuss the datasets chosen for this analysis in section \ref{sec.data}. We will then in detail discuss the profiling method applied, pre-subtractions and post-processing, and the automated wind estimation techniques in section \ref{sec.method}. We discuss the generation and injection of simulated atmospheric layers in section \ref{sec.inj} and the resulting successful recovery in \ref{sec.inj_rec}. The found wind velocities are detailed in section \ref{sec.layer_detect} and then compared to other measurements on the summit in section \ref{sec.disc}. Finally, we will conclude with method limitations and future applications of this method, section \ref{sec.conclusion}.

\section{Data Collection} \label{sec.data}

    
    \subsection{Selected 'imaka Data}\label{sec.runs}
    
    Data were collected from 2018 to 2021 in a total of 12 runs on the 'imaka instrument. From those runs, three were selected, based on quality of open loop telemetry, features in the data, and clarity of correlation peaks. Summary of files chosen can be seen in Table \ref{table:data}. During these runs telemetry data, consisting of WFS pixel values, calculated subaperature centroids, DM voltages, and system configurations were collected from up to 5 simultaneously running 8x8 subaperture Shack-Hartman WFS. For our work we focused on the open loop data sets, where turbulence information is not affected by AO corrections. Though open loop telemetry can be reconstructed from closed loop data, and has been show in other papers, the closed loop telemetry files available will be added in a later study.
    
    
    \begin{table}
    \centering
    \begin{tabular}{|l|l|l|l|l|l|} 
    \toprule
    \hline
     Target             & Date     & \# of total files & WFS used & integration time & Int t sum  \\ 
    \hline
    \multirow{2}{*}{HD184336}   & 2018 05 31 & 25                & 1-3      & 806.36           &            \\ 
                             & 2018 06 01 & 69                & 1-3      & 2225.60          & 3,031      \\ 
    \hline
    \multirow{4}{*}{Orion2} & 2018 12 19 & 110               & 1-5      & 1637.23          &            \\ 
                             & 2018 12 21 & 97                & 1-3, 5   & 1854.69          &            \\ 
                             & 2018 12 22 & 80                & 1-3, 5   & 1749.85          &            \\ 
                             & 2018 12 23 & 144               & 1-3, 5   & 5708.29          & 10,950     \\ 
    \hline
    \multirow{4}{*}{HD184336}   & 2021 04 29 & 25                & 1-3      & 2115.66          &            \\ 
                         & 2021 04 30 & 19                & 1-3      & 2651.04          &            \\ 
                         & 2021 05 01 & 16                & 1-3      & 2209.45          &            \\ 
                         & 2021 05 02 & 37                & 1-3      & 3010.96          & 9,987      \\ 
    \hline
    \textbf{Total} &          & \textbf{622}      &          &                  & \textbf{23,968s}    \\
    \hline
    \end{tabular}
     
    \caption{Table shows all targets, dates, and number of circular buffers used. Note that the 12th run had longer exposures, so even with fewer files per night, still has an equivalent integration time on camera. Different number of WFSs are used due to the number of NGS available per constellation. }
    \label{table:data}
    \end{table}
    
    \subsection{External Data}\label{sec.ext_data}
    
    Data for ground layer wind velocities are taken from the Canada-France-Hawaii's weather station, which acquires wind speed and direction at the telescope dome roughly once a minute. The CFHT weather station is located about 200 meters North of the UH88 between the CFHT and Gemini Observatory enclosures. We use this data as a reference for the speed and direction of lower turbulence layers.

    \begin{figure}[!htb]
   \begin{minipage}{0.48\textwidth}
     \centering
     \includegraphics[width=.98\linewidth]{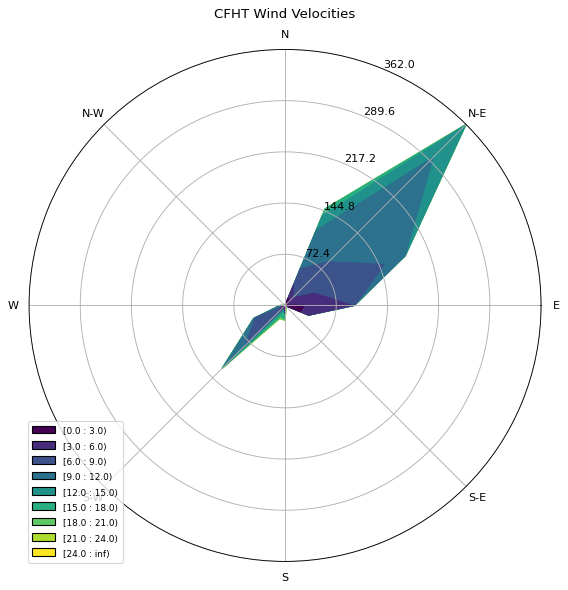}
     \caption{CFHT weather station wind velocities for observed times. Shows count by wind declination and speed. This ground layer data shows a majority of our data is expected in the NE direction. Speed in m/s.}\label{fig.wind_cfht}
   \end{minipage}\hfill
   \begin{minipage}{0.48\textwidth}
     \centering
     \includegraphics[width=.98\linewidth]{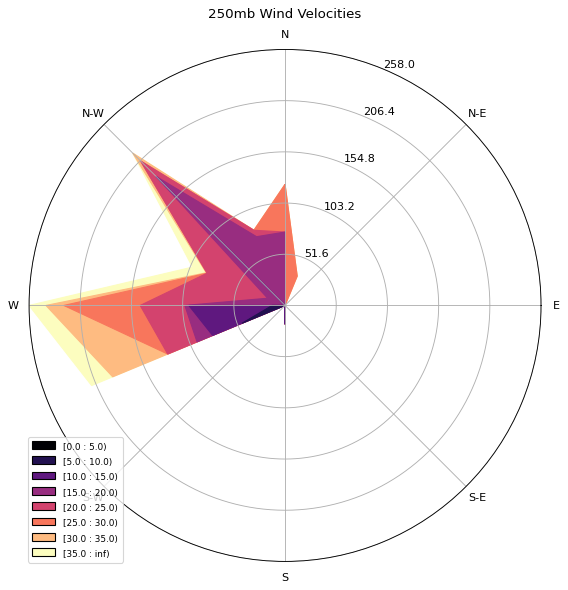}
     \caption{250mb wind velocity model. This upper layer estimate shows a majority of our data is expected in the N and W directions, and on average has faster winds than lower layers. Speed in m/s.}\label{fig.wind_250}
   \end{minipage}
\end{figure}

    Data for upper layer wind speeds are taken from the Global Forecast System (GFS) weather forecast model produced by the National Centers for Environmental Prediction. The GFS upper atmosphere (250mb) wind model predictions "nowcasted" at 12hr UT above Maunakea are used as a nightly proxy for the upper winds. The 250mb layer corresponds roughly to 10km above sea level (~6km above the site). A summary of the wind speeds and directions for ground and upper atmosphere layers for the nights observed are summarized in figures \ref{fig.wind_cfht} and \ref{fig.wind_250}.

\section{Method} \label{sec.method}

    \subsection{Pre-Subtractions} \label{sec.pre_sub}
    
    Our data, for each of up to 5 WFSs and for x- and y-slopes independently, has two types of aberrations subtracted before the correlation is applied.  The first subtraction is a static slope vector map from each instantaneous slope map. This is done by calculating an average slope vector map over the full data set, where each dataset typically contains 30-150 seconds worth of data. The static slope map is removed to remove static spatial correlations which can interfere with the detection of moving covariance peaks. However this also removes correlations for turbulence that varies over longer timescales (e.g. dome seeing). We then subtract off the instantaneous global tip tilt, found by taking the mean slope for valid subapertures at each exposure. Subtracting global tip and tilt mitigates errors from telescope tracking, but removes information about the correlations in the global tip/tilt.  This ordering was specifically chosen to remove lingering statics. 


    
    \subsection{Correlations} \label{sec.corr}
    
    Once cleaned, each circular buffer's active wavefront sensors were correlated with themselves, (auto-covariance) and with all other wavefront sensors (cross-covariances). As a layer of turbulence crosses over the telescope's line of sight, one could assume, via the frozen flow hypothesis, that its detailed shape remains constant other than the simple translation across the pupil. This means at each point in time after an initial wavefront measurement, we see a near replica of that initial wavefront but slightly shifted. A correlation of these two wavefronts will show a peak at the shift where the wavefronts are most similar. Here we can use the correlation of a WFS's slopes at a time shift to see the distance tracked by the covariance peak \cite{wilson2002SLODAR}: 

    \begin{equation}
    A(\delta i, \delta j, \delta t)=\frac{\left\langle\sum_{i j} s_{i j}(t) s_{i+\delta i, j+\delta j}(t+\delta t)\right\rangle}{O(\delta i, \delta j)}
    \label{e.autocorr}
    \end{equation}

    Each element of the auto covariance matrix, in three dimensions, is defined by taking the time averaged $\left\langle \right\rangle$ summation over one WFS slope indices $\sum_{i j}$ where each is the element wise multiplication of the slope matrix $ s_{i j}(t)$ with the spatially and temporally offset slope matrix $s_{i+\delta i j+\delta j}(t+\delta t)$. These values are normalized by how many total elements were overlapped in the summation calculation $O(\delta i, \delta j)$.

    Averaging over valid WFS, the auto-covariance output is dominated by a covariance peak that does not move in time. Namely we see a dominant peak at the center of the covariance map that lasts for large separations in time in the temporal correlation. We postulate that this is a signature of turbulence, or seeing, that arises inside the telescope enclosure (e.g. "dome seeing") as static aberrations from other sources were previously subtracted. This signature is discussed in the results section. We note that because the dome seeing is normally so strong, its covariance peak dominates the peaks from all other layers. This hampers the identification of moving layers. Applying either a rolling average of the covariance matrix or the average covariance map shape uncovers the next most dominant turbulence layer. 
    
    Though auto-covariances of individual WFSs can provide information about the strongest layers, they do not provide information on the altitudes of the observed peak(s). We can use the arrangement of natural guide stars and the correlation between wavefront sensors (e.g. cross-covariance) to determine the height of a turbulent layers by seeing where the peak is located in the cross-covariance between two different WFSs at $\delta t=0$. Just as we see wavefront patterns repeat in time, they will repeat in space depending on the angular separation of targets. Here, we denote that the cross-covariance uses two separate WFSs, $s$ and $s^{\prime}$, the only deviation of eq. \ref{ccor} from \ref{e.autocorr}.

    \begin{equation}
    C(\delta i, \delta j, \delta t)=\frac{\left\langle\sum_{i j} s_{i j}(t) s_{i+\delta i j+\delta j}^{\prime}(t+\delta t)\right\rangle}{O(\delta i, \delta j)}
     \label{ccor}
    \end{equation}
    
    Once all wave fronts have been auto and cross-correlated, we average the x and y slopes, giving one cube per WFS x WFS permutation. 
    
    \subsection{Post-Processing} \label{sec.post_sub}
    
\begin{figure}[ht]
    \centering
    \includegraphics[width=0.88\textwidth]{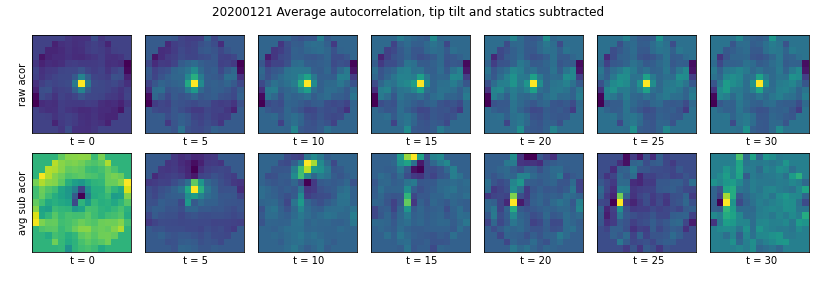}
    \caption{(First row) Dome seeing dominates the raw auto-covariance matrix. (Second row) An upward traveling peak is clearly visible once a background average is subtracted. Scaling is in arbitrary units, t is in units of exposure frames.}
    \label{fig.acorr_ex}
\end{figure}

\begin{figure}[ht]
    \centering
    \includegraphics[width=0.88\textwidth]{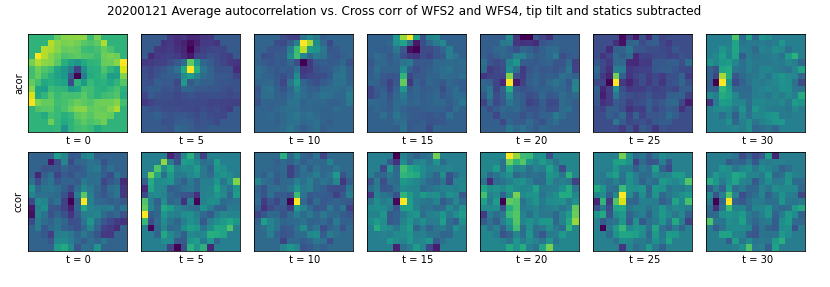}
    \caption{(First row) Auto-covariance shows two peaks, one moving upward quickly, and another moving to the left. (Second row) Cross-covariance shows only one peak, moving to the left. The t=0 frame can be used to derive altitude from separation of peak from center.}
    \label{fig.ccorr_ex}
\end{figure}

    After correlation, there are still noticeable artifacts to each covariance map. Most prominent is a strong central peak, which has been attributed to dome seeing\cite{guesalaga2014cn2}, a source of quasi-stationary turbulence. In order to better detect and extract the other layers, we experimented with various subtractions post correlation. The average of the first $n$ time steps creates a background map that is subtracted from the covariance map. Short windows, where n=50 time steps are used in the background subtraction, suppress slower moving peaks but accentuate faster layers. A longer window of n=200 takes out the largest, longest peak with minimal suppression of slow but moving features. The impact of these post-subtractions can be seen in figure \ref{fig.acorr_ex}. The difference between auto- and cross-covariances with backgrounds subtracted can be seen in figure  \ref{fig.ccorr_ex}.
    
    \subsection{Peak Extraction and Wind Estimations} \label{sec.estimate}
    \label{sec.peak_est}
    
    Most, if not all of the peaks found in our covariances are detectable by eye. Given the number of files available for processing, an automated system for peak extraction was deemed necessary. This method finds peaks from the average of the x-slope and y-slope covariance maps. 
    
    Our layer detection algorithm is based on the assumptions that (1) ground layer and free-atmosphere layers are seen in the auto-covariances, (2) that a layer can be identified as a ground-layer by being in both the auto-covariances and the cross-covariances, and (3) that a ground layer is at or very close to the ground.  The first two assumptions are true up to our detection sensitivity.  We discuss our sensitivity limits in sections \ref{sec.inj} and \ref{sec.inj_rec}. The last assumption is not guaranteed to be always valid even within our detection sensitivity limit.   In particular, layers between about 100m and 600m appear in the cross-covariances to originate away from the center of the covariance map.   Since our current algorithm only looks for covariance peaks that start near the center of the correlation function we misidentify these layers as FA layers since they are detected in the auto-covariance but not the cross-covariance.  We see such layers in a small fraction of on-sky data but the majority of ground layers are seen, at least by eye, starting within a resolution element of h=0.  We plan to generalize the algorithm but have not done so yet.  
    
    Finally we note that our ability to detect layers is limited by the signal-to-noise ratio of the observations and by the spatial and temporal resolution of our data.  For this approach the first sets a limit on the strength of the layer we can detect while the resolution of the data limits how well we can resolve layers that move with similar velocities.  We note that given the our aperture size and our sampling rate we only obtain a handful of overlapping wavefront measurements for a layer moving at 50 m/s before it exits the pupil. This is an extreme wind speed for the ground but can exist when the jetstream is overhead.  That is generally not the case for Maunakea. On the other extreme, we are limited to the slowest layer we can detect by the length of our telemetry data.  These are generally 4000-27000 steps of the AO system and we generate a temporal cross-covariance map up to time separations of 1000 steps (so at 180Hz this is 5.5 seconds). The layer detection algorithm is as follows:
    
    
    First we calculate a detection level for each element in the temporal covariance (e.g. 15x15x1000 elements).  For each time step, we step out from the center pixel and calculate a mean and standard deviation for the elements within an annulus centered on the center pixel. For each element in that time step we then define the detection level as is deviation in standard deviations from the mean.   We do this for all annuli for a particular time step and then repeat for all time steps.  In the end we have a deviation from the mean value for every element in the temporal covariance map.
    
    Second, a detection is set to some threshold (typically three standard deviations above the mean), so that any detection above this threshold in our detection cube is identified as an individual covariance peak detection. For each covariance peak detection, we then calculate a velocity associated with this peak.  Namely, we convert its location in time and radius to a velocity (speed and direction). For this method, we assume that peaks start at center, which is always the case for auto covariances, but for cross covariances this assumption ignores the non zero ground layer peaks where starting off-center indicates altitude. A speed is determined from the (x,y) location in the covariance and the timestep. Direction can be calculated in degrees from North, but a projection onto the unit circle (e.g. an "exit" coordinate on the unit circle) is found to be a better representation for clustering of the data. This step then concludes with a list of speeds and "exit" (x,y) coordinate) of each detected covariance peaks along with their detection significance in standard deviations.
    
    Finally, since we track covariance peaks though the temporal correlation sequence, the list of detected peak track the motion of a covariance peak through the time axis of the cube.  We apply a k-means clustering algorithm (specifically the CLUSTER algorithm built into IDL) to the detected speed and the direction (e.g. the (x,y) exit coordinates). For each cluster we calculate the mean speed and direction (degrees East of North) and the mean of the individual covariance peak detections in that cluster.
 
    \subsection{Simulation Injections} \label{sec.inj}
   
    There are a few parameters that can be tuned in this implementation of the wind profiling method. For example the covariance peak detection threshold, the type of clustering, the type of background covariance removed in the covariance maps, and the length of the covariance file. To give us some insight into how these parameters effect our ability to recover layers of different speeds and optical turbulence strength we injected simulated layers of known strength and velocity into a set of representative telemetry data. From these experiments, we established that a threshold of 3 standard deviations, with a mean background subtraction and a covariance file length of 50 or 200 time steps led to a reasonable recovery rate.
    

    Injected layers were generated from a base simulated phase screen. The initial screen was 30 meters square, the turbulence had Von Karman statistics with $r_{0,sim}=0.29$ m and $L_0=30$ m. This screen was then extruded to a length of 2150 m along one axis, the length needed for replay in our simulations at the maximum wind speed used in our tests. Layers of different strengths,  $r_{0,inj}$, were generated by scaling the phase values by $(r_{0,inj}/r_{0,sim})^{-5/6}$.  The AO simulation contained a single free atmosphere layer at an altitude of 7km above the site and a set of four guide stars/wavefront sensors on a square asterism of guide stars with a similar size as the on-sky asterisms.  The simulations do not match the on-sky data but we note that at the guide star separations there are several meta-pupils of separation between each wavefront sensor measurement so we expect that correlations in the free-atmosphere are well below our detection sensitivities. The number of guide stars/wavefront sensors in the on-sky telemetry data varies from 3-5 but the analysis in this section, we only consider the first three wavefront sensors in the simulated and on-sky telemetry data.  

    Ten on-sky telemetry files were chosen to be representative of the overall data set.  One data set was chosen per night and they have a mean standard deviation in the annuli of the covariance maps that is typical of the noise from the observations for that night.  The ten chosen data sets typically have at least one atmospheric layer, seen by eye, in the covariance maps.   The speed and direction are a reasonable representation of the overall data set since the layer velocities and directions tend to stay fairly constant throughout a given night.
        
    A single simulated free-atmosphere layer (or rather set of slopes) with a known r0, speed, and direction were injected into each of these representative on-sky data sets.  We generated multiple "injection" data sets by varying the injected layer's r0, speed, and direction. The layer strengths chosen (0.10 to 0.50 in increments of 0.05), in $r_{0}$ seeing, were selected to span very strong peaks at 0.10 to faint seeing at 0.50. The speeds chosen (1, 3, 6, 12, 24, 36m/s) were picked to span from the lowest limits of detected speeds where noise usually contaminates files to the high end of meteorological measured speeds for upper atmosphere layers (refer back to Figure \ref{fig.wind_250}).  The layer directions were limited to the four cardinal directions. 
        

\section{Analysis} \label{sec.analysis}
    \subsection{Injection Recovery} \label{sec.inj_rec}
    
\begin{figure}[ht]
        \centering
        \includegraphics[width=0.95\textwidth]{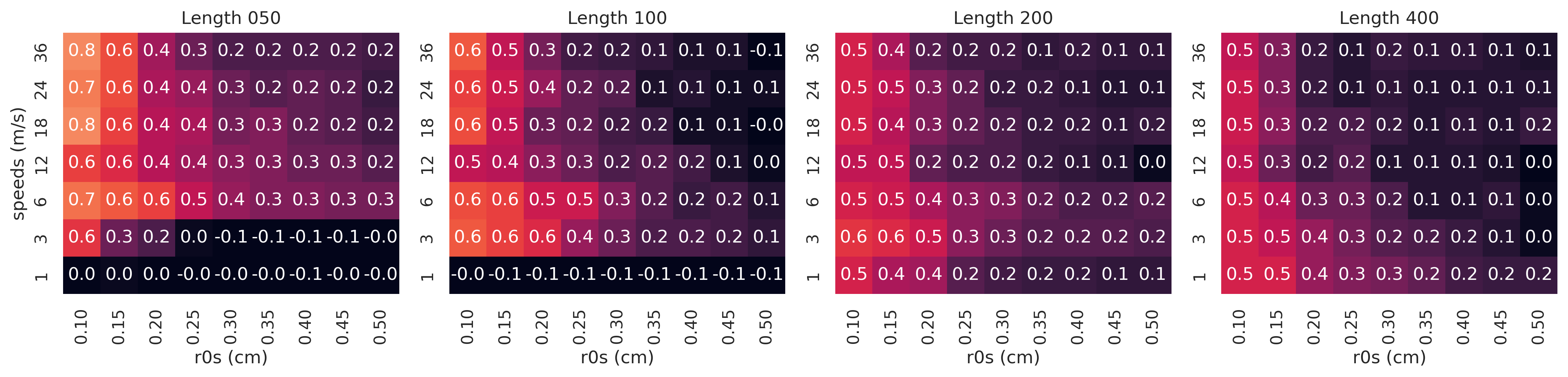}
        \caption{The successful recovery of injected layers across varying turbulence strengths and layer speeds. We see variation of successful detections across the depicted covariance lengths. The method is much more sensitive to stronger seeing.}
        \label{fig.inj_rec}
    \end{figure}

    The recovery of injected layers at varying speeds, cardinal directions, and strengths gives perspective on two main questions on our analysis: (1) the method's expected recovery rates and (2) the range of speeds and strengths is this method sensitive to on the UH 2.2m. Figure \ref{fig.inj_rec} depicts the successful recovery of an injected layer of strength r0 and speed, in four cardinal directions, into ten on-sky files. A successful recovery is determined by a returned layer that is within range of the injected peak. To ensure that the true peaks aren't detected instead of the injected, we run this process on both plain and injected circular buffers, and subtract the on-sky files success rates as a control. This gives a percentage of successful detection above the on-sky detection baseline. 
    
    \subsubsection{Estimation parameters}
    Questions we had for our estimator before beginning the injection process were centered around the optimum length of post-subtraction and detection thresholds. From these experiments a detection threshold of 3 standard deviations from the mean was determined as it returned the optimal number and accuracy of detections. The background subtraction length is not as obvious. Figure \ref{fig.inj_rec} illustrates the split between the recovered speed of a short subtraction an a longer subtraction, where for a shorter length, no slow speeds are recovered, where longer subtractions have greater difficulty at higher speeds. All subtraction lengths struggle at recovering fast but weaker layers. 
    
    \subsubsection{Estimation limitations}
    In section \ref{sec.peak_est} a preliminary prediction of sensitivities were made from the lengths of our AO data and the rate of telemetry acquisition. The injection and extraction of peaks from on sky noise gives a experimental nuance to those estimation limitations by understanding the intersection of strengths and speeds in detection. As seen in figure \ref{fig.inj_rec}, we can see in general that any stronger layer is more likely to be detected than weaker layers, regardless of speed or covariance length. We are able to detect layers as slow a 1m/s out to our limit of 36 m/s at similar rates, given the length of the background subtraction.
    
    
    

    \subsection{Layer Detections} \label{sec.layer_detect}
    
\begin{figure}[ht]
    \centering
    \includegraphics[width=0.88\textwidth]{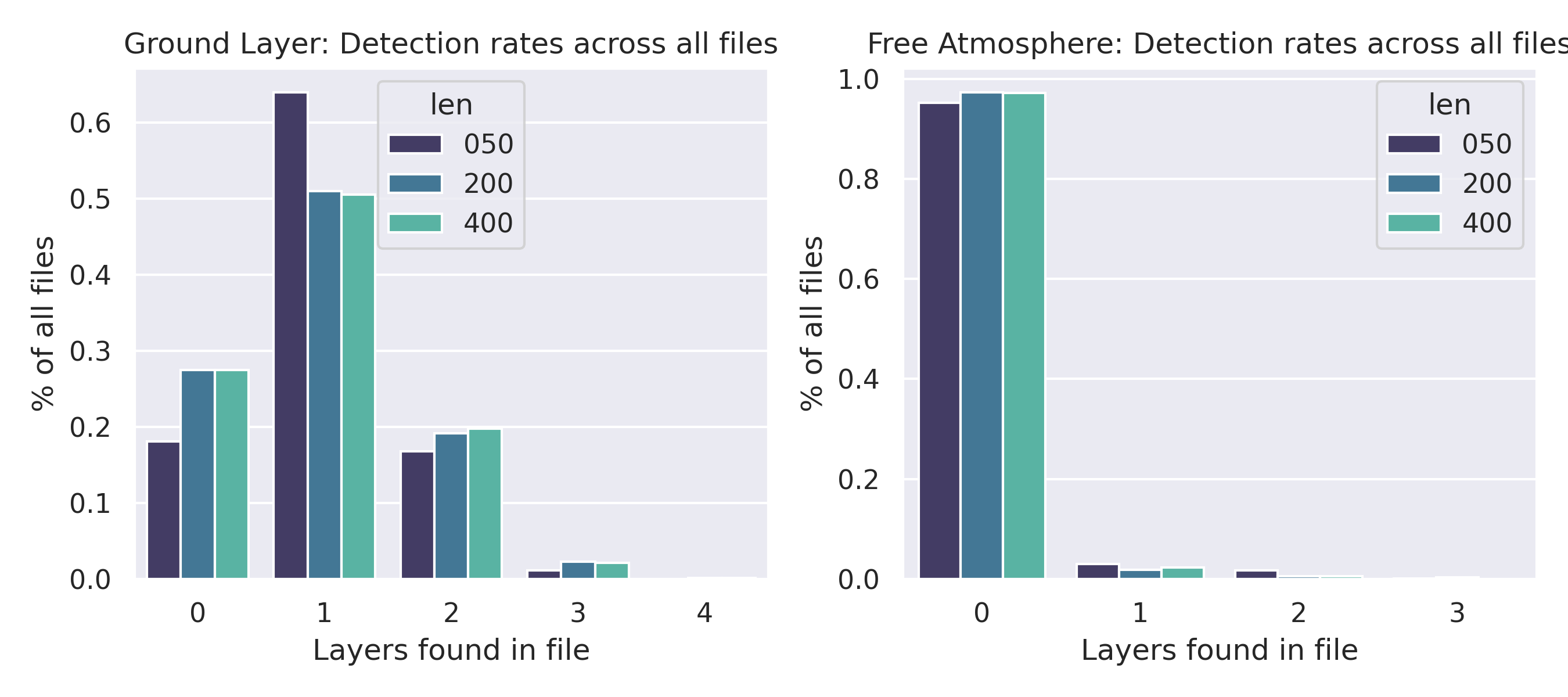}
    \caption{The number of layers across our files chosen. A majority of files have a least one ground laye (GL) identified by our algorithm. Only a small fraction of files have free atmosphere (FA) layers identified. Files with three or more GLs identified likely have a peak miss-identified, either noise or FA. }
    \label{fig.detection_rate}
    \end{figure}

\begin{figure}[ht]
    \centering
    \includegraphics[width=0.95\textwidth]{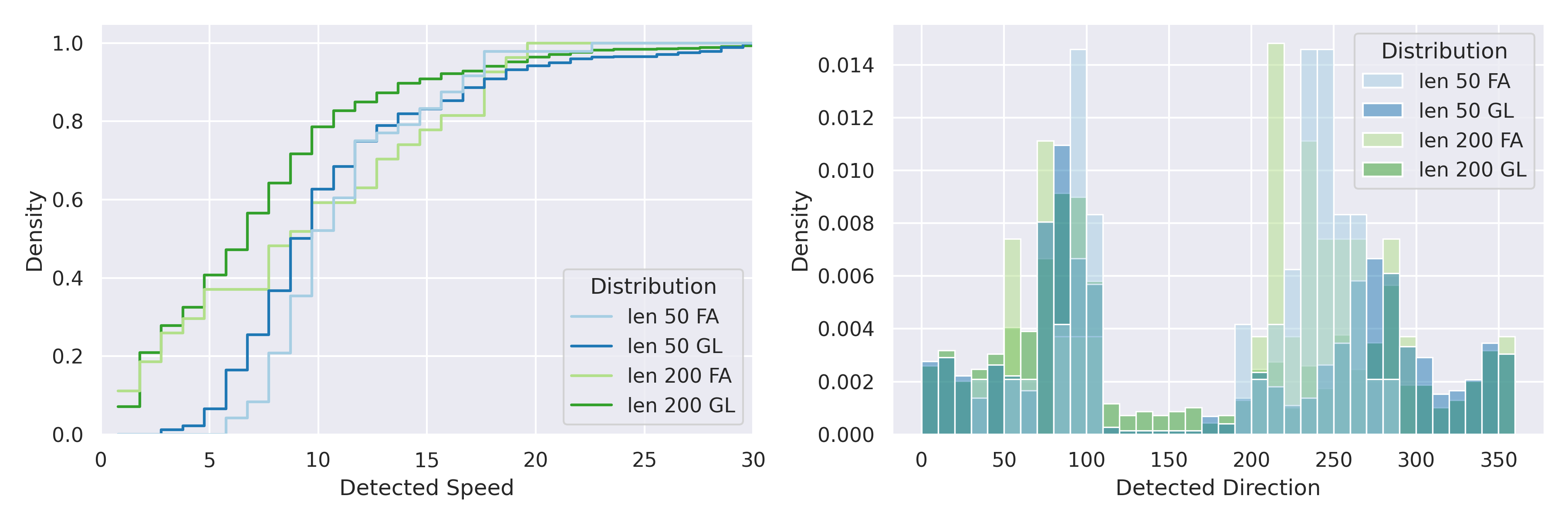}
    \caption{The distribution of speeds and directions across all found layers. A shorter length of background subtraction of 50 cuts out slower layer speeds, but increases detection of faster FA layers. The speed distributions look roughly similar between FA and GL, with a slight bias towards faster layers. There is a visible shift in layer directions between ground and upper atmosphere, as expected.}
    \label{fig.detection_cdf}
    \end{figure}

    Next, the estimator is run on telemetry files and we report on how many layers and of which types were found in each circular buffer in Figure \ref{fig.detection_rate}.
    The distribution of speeds and directions detected in these models can be seen in Figure \ref{fig.detection_cdf}. Note that files can have both free and ground layers detected, both, or neither. This gives a rough overview of the detected values and are further compared and contextualized with external data in section \ref{sec.disc}.
    
    At least one ground layer is found in 70-80\% of the telemetry files examined, while FA layers are detected in less than 10\% of files. In other words, about 300 files had only one ground layer detected, while 100 or so files had at least two or more ground layers found. The relationship between the two or more layers is more dependent on the window length, indicating that these detections are likely noise or miss-labeled free atmosphere detections. While there is usually a strong ground layer, there isn't as often a detectable upper atmospheric layer. The fast, fainter layers fall into a blind spot seen in our injection simulations. Future work will need to increase sensitivity to FA layers. 
    
    
    These plots are shown with multiple subtraction lengths to illustrate how the post-correlated subtractions bias the resulting detected layers. In Figure \ref{fig.detection_rate} there is a noticeable difference between shorter and longer lengths in how many files are found, though less difference between length 200 and 400 subtractions. In Figure \ref{fig.detection_cdf} the len 50 subtraction is show to under-detect slow layers, as predicted. When the window used for the average subtraction is shorter, it averages out these lower moving peaks gradual motion. However, we also can tell that the faster moving peaks, especially in the FA layer, are under-detected with a slower moving subtraction. A cut at 5m/s is taken to combine the results of the two methods.

\section{Discussion}\label{sec.disc}

    
    
    
    
    \subsection {Ground layers and CFHT and GFS}\label{sec.comp_gl}
    
\begin{figure}[ht]
    \centering
    \includegraphics[width=0.88\textwidth]{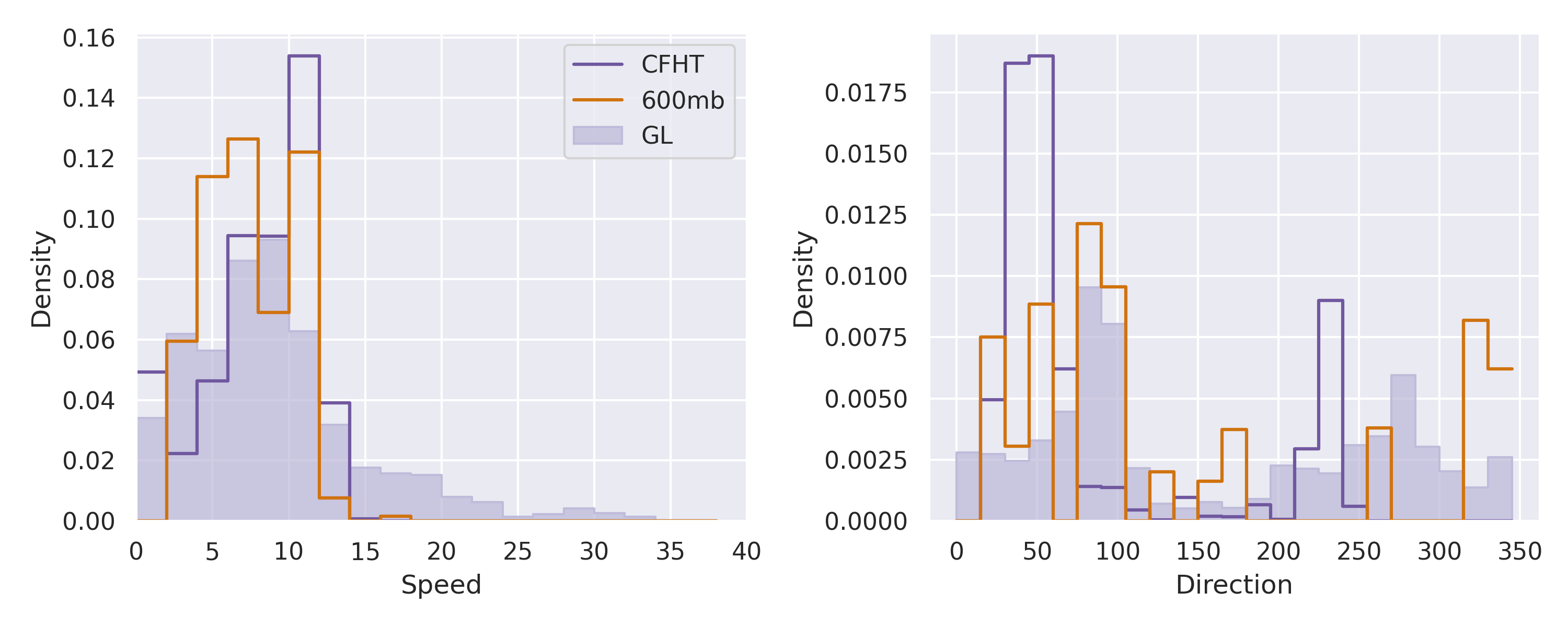}
    \caption{Comparison the the CFHT weather tower detections. Speeds comparison shows a wide spread in speeds, including a section of faster layers underestimated by our algorithm. Directions are also widely spread. The subsection}
    \label{fig.CHFT_comp}
    \end{figure}

    There are two external comparison points for ground layer turbulence. The first is from the CFHT weather station, and the second is the GFS model prediction at the pressure of the summit, around 615mb. Figure \ref{fig.CHFT_comp} shows the distribution of detected ground layers (GL) compared to the temporally closest datapoint collected for each found layer. 
    
    The speed distribution shows a tail of higher speed detections than either of the external models. There is also an over representation of slower speeds. The direction distribution shows a much wider spread, likely due to local turbulence at both the telescope domes, and near the CFHT. Interestingly, the two main peaks of the found distribution are around 45 degrees off from the CFHT distribution. This offset seems to be from the last 5 years, looking at historical trends of wind directions captured by the weather station. 
    
    \subsection {Free atmosphere layers and GFS models}\label{sec.comp_fa}
    
\begin{figure}[ht]
    \centering
    \includegraphics[width=0.88\textwidth]{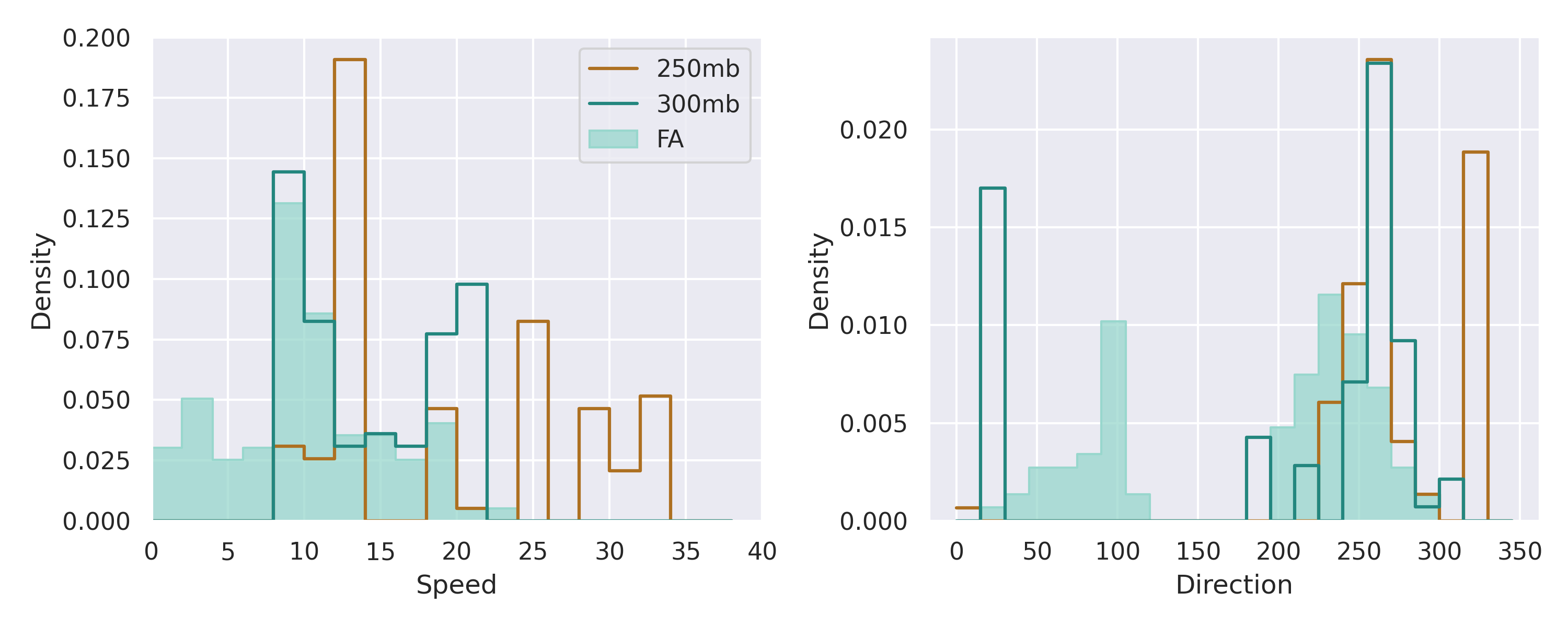}
    \caption{Comparison of detected free atmosphere peaks with two pressures closest in altitude to expected layer location.}
    \label{fig.GFS_comp}
    \end{figure}

    There are noticeably fewer free atmosphere than ground layer peaks detected, and so the potential analysis and conclusions here will limited. We see in figure \ref{fig.GFS_comp} that our method once again over reports on slower peaks. Of the two proposed atmospheric pressures, they have similar direction distributions, but the 250mb has on average faster moving peaks. Even though our detected FA layers seem to align closer to the 300mb distribution, agian this is where our method has been shown to under-perform. 
    
\section{Conclusion} \label{sec.conclusion}

    The unique arrangement of the 'imaka GLAO experiment allowed this project to use its multiple guide stars and open loop telemetry to detect turbulence speeds through a wind profiling technique. We have used correlation and peak extraction to examine hundreds of open loop telemetry files from the UH2.2m telescope. A ground layer atmospheric layer was identified in 80\% of the files of interest, while free atmosphere layers were detected in less than 10\%. The detected distributions are seen to follow external meteorological measurements with some outliers. Injection of simulated layers depicts the method's sensitivity, which covers the range of speeds typically measured at Maunakea ($1-36$m/s) but at the smallest $r_0$s from 0.1-0.25.
    
    Future work on this project could increase the sensitivity to free atmospheric layers by using more sensitive subtraction techniques, as these significantly change   distributions of detections. The detection algorithm used here has room for improvement, such as including off-center peaks and accounting for the detection of secondary peaks. With improved identification of FA layers, this technique could be another tool in filtering algorithms for wide-field ground layer AO observations.

\acknowledgments     

    We acknowledge the NSF award 1910552 that funded this project. We would also like to thank Ryan Lyman for archiving and providing the CFHT and GFS weather data and for recommending sounding data (not used here). This work is based on observations made by University of Hawaii 2.2-meter (88-inch) telescope. We wish to extend our special thanks to those of Hawaiian ancestry on whose sacred mountain of Maunakea, we are privileged to be guests. Without their generous hospitality, the observations presented herein would not have been possible.

\bibliography{ref} 
\bibliographystyle{spiebib} 


\end{document}